\documentclass[prd,showpacs,amsmath,twocolumn,floatfix,amssymb, preprintnumbers, nofootinbib, superscriptaddress]{revtex4} 
\usepackage{epsfig,dcolumn}
\usepackage{graphicx}
\usepackage{comment} 
\usepackage{verbatim}
\usepackage[usenames]{color}
\usepackage{graphicx}
\usepackage{bm}
 \usepackage{ifpdf}

\def\lsim{\mathrel{\rlap{\lower4pt\hbox{\hskip1pt$\sim$}}
    \raise1pt\hbox{$<$}}}
\def\gsim{\mathrel{\rlap{\lower4pt\hbox{\hskip1pt$\sim$}}
    \raise1pt\hbox{$>$}}}
\begin{document}

\title{ Analytic continuation of Pasquier inversion representation of Khuri-Treiman equation }

\author{Peng~Guo}
\email{pguo@jlab.org}
\affiliation{Physics Department, Indiana University, Bloomington, IN 47405, USA}
\affiliation{Center For Exploration  of Energy and Matter, Indiana University, Bloomington, IN 47408, USA.}

%\preprint{JLAB-THY-}

\date{\today}

\begin{abstract} 
The single integral form   of Pasquier inversion representation of Khuri-Treiman (KT) equation presents great advantages for describing final state interaction of three-body decay or production processes. However, the original form of Pasquier inversion representation is only given in decay region and   regions below.  For the  regions above,  analytic continuation  of original form is required. Because of non-trivial nature of analytic continuation procedure, it is the purpose of this work to obtain a well-defined   Pasquier inversion representation of KT equation for all the energy range.
  \end{abstract} 

\pacs{ }

\maketitle

%%%%%%%%%
%  Introduction  %
%%%%%%%%%

\section{Introduction}
\label{intro} 
The theoretical framework for describing low energy  hadronic three-body interaction has attracted  significant attentions in the past, different approaches have been developed, such as field theory based    Faddeev   and Bethe-Salpeter type equations \cite{Faddeev:1960su,Faddeev:1965su,Taylor:1966zza,Basdevant:1966zzb,Gross:1982ny}, and dispersion relation orientated  Khuri-Treiman (KT) equation \cite{Khuri:1960zz,Bronzan:1963xn}. In   processes,  such as $\eta \rightarrow 3 \pi$,    three-body final state interaction has been reported to play a important role in explaining the discrepancy  of Dalitz plot expansion parameters between experimental measurements and theoretical calculations  \cite{Gasser:1984pr,Kambor:1995yc,Anisovich:1996tx,Bijnens:2002qy,Bijnens:2007pr,Colangelo:2009db,Zdrahal:2009cp,Schneider:2010hs}. 

Among different methods,  dispersion approach based KT equation shows some advantages because  of it's simplicity of formalism  and analogue to naive isobar model approximation \cite{Goradia:1975ec,Ascoli:1975mn}. Since  first proposed  
in \cite{Khuri:1960zz},  KT equation has been further developed   by many authors \cite{Bronzan:1963xn,Aitchison:1965kt,Aitchison:1965zz,Aitchison:1966kt,Pasquier:1968zz,Pasquier:1969dt,Guo:2014vya}. Original form of KT equation is written in a form of  double integrals dispersion equation, one integral comes from the dispersion integration and another is related to   partial wave projection. By using Pasquier inversion technique \cite{Pasquier:1968zz,Aitchison:1978pw,Guo:2014vya}, the order of two integrals can be exchanged, and it results in a single integral representation of KT equation that is more suitable for numerical computation \cite{Aitchison:1965kt,Aitchison:1965zz,Aitchison:1966kt,Pasquier:1968zz,Pasquier:1969dt,Guo:2014vya}. Unfortunately, original form of Pasquier inversion representation of KT equation  is not well-defined  for all the energy range, in fact, the original form is only given in the physical decay region and    regions below. For other energy regions, analytic continuation of  Pasquier inversion representation of KT equation has to be carried out deliberately to avoid   singularities generated by contour integrations. As will be discussed in this work, the energy range above two-particle threshold is divided  by a complex contour   into three parts: decay, unphysical and scattering regions. Unphysical region is disconnected from decay and scattering regions,
in this region, original form of KT equation has to be modified and an extra term is needed to keep solution of KT equation staying on physical sheet. Due to  non-trivial procedure of analytic continuation,  we describe  some details of analytic continuation in this work, and present   a well-defined form of   Pasquier inversion representation  of KT equation  in all energy regions.

The paper is organized as follows. The original form of Pasquier inversion representation    of KT equation  is briefly introduced in Section \ref{formalism}.  The  procedure of analytic continuation  is described in Section \ref{analyticpasquier}. The summary and conclusion are given in Section \ref{summary}.

\section{Subenergy dispersion approach to three-body final state interaction}\label{formalism}
A general   amplitude for a particle with spin-$J$ decays into three spinless particles, such as in  $J/\psi $ decays  \cite{Guo:2010gx,Guo:2011aa}, reads
\begin{equation}
\langle 1 2 3 , \mbox{out}| J(\lambda) , \mbox{in} \rangle= i (2\pi)^{4} \delta^{4} (\sum_{i=1,2,3} p_{i}-P) T_{\lambda},
\end{equation}
where we denote the four momenta by $p_{i},P$ for i-th final state particle and initial decay particle, and $\lambda$ is the spin projection of the initial state along a fixed  axis.  Suppressing  the isospin coupling among initial and final states,   the  amplitude $T_{\lambda}$ is given by, 
\begin{align}\label{decayamp}
T_{\lambda}(s,t,u)= &     \sum_{S,L,\mu}  N_{SL \mu} D^{J *}_{\lambda, \mu} (r_{s})     d^{S}_{\mu,0}(\theta_{s}) a^{(s)}_{SL} (s)\nonumber \\
& + (s \rightarrow t) + (s \rightarrow u),
\end{align}
where   the invariants are defined by \mbox{$s =(p_{1}+ p_{2})^{2}$}, \mbox{$t =(p_{2}+ p_{3})^{2}$} and \mbox{$u =(p_{3}+ p_{1})^{2}$}, and they are constrained by relation, \mbox{$s+t + u = M^{2} +\sum_{i} m_{i}^{2}$} ($m_{i}$'s are final state particles masses and $M$ is mass of initial particle),  
 and \mbox{$N_{SL \mu} =  \sqrt{(2S+1) (2L+1)} \langle S\mu ; L0 | J \mu \rangle$}.   The spin of pair $(12)$ is  denoted by $S$, and the relative orbital angular momentum between $(12)$ and the third particle is given by $L$.  $  \theta_{s}$ is polar angle of particle-1 in    pair $(12)$ rest frame. The rotation $r_{s}$, which is given by three Euler angles \cite{Guo:2010gx,Guo:2011aa},   rotates the standard configuration in $(12)3$ coupling scheme to the actual one. In the standard configuration of $(12)3$ coupling (rest frame of three-particle), third particle moves along negative $z$ axis while particle-1 and -2 move in the $xz$ plane. The amplitudes in $(23)1$ and $(31)2$ coupling schemes (denoted by $t$- and $u$-channel respectively) are defined in a similar way as in $(12)3$ coupling (denoted by $s$-channel).
   The dynamics of decay  process  are described by   scalar functions $a^{(s,t,u)}_{SL} $,  which only depend on subenergy of isobar pairs and possess  only unitarity cut in  subenergy by assumption \cite{Aitchison:1965zz,Aitchison:1966kt,Pasquier:1968zz}. 
 
 For   simplicity, in the following discussion, we  consider  the decay of  a scalar particle, \mbox{$J=0$}, and    truncate the partial waves to include only $S$-wave: \mbox{$S=L=0$}.  Masses   of final particles are assumed identical:  \mbox{$m_{1}=m_{2}=m_{3}=m$}, and    sub-channels are assumed symmetric: \mbox{$a_{00}^{(s)}=a_{00}^{(t)}=a_{00}^{(u)}=a$}. Thus, the decay amplitude  is simply given by sum of three terms,
\begin{align}\label{decayampswave}
T(s, t, u) = & a (s) + a(t) + a(u)   .
\end{align}

\subsection{Khuri-Treiman equation and Pasquier inversion representation  }
\begin{figure}
\includegraphics[width=0.48\textwidth]{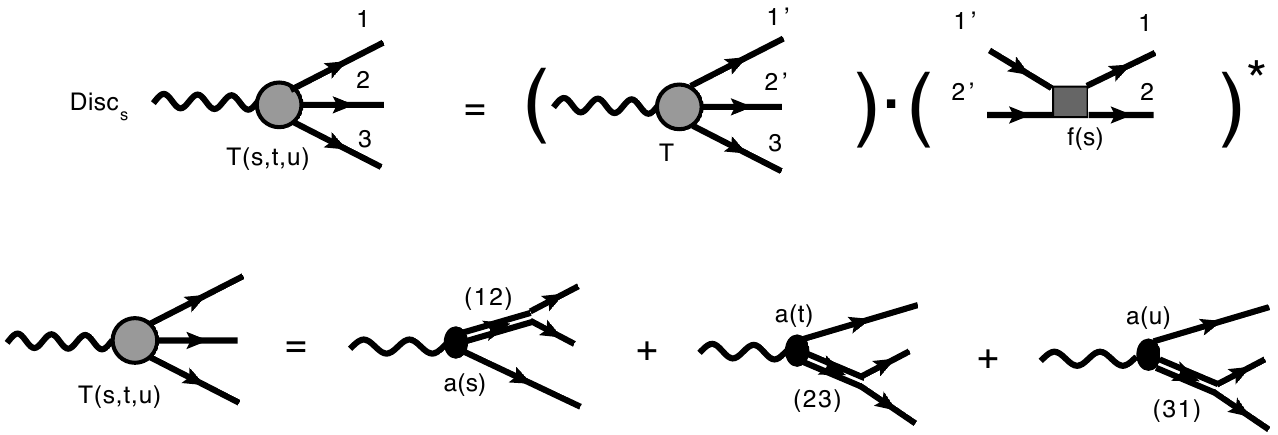}
\caption{    A diagrammatic representation of discontinuity relations in   Eq.(\ref{fdis1}). 
\label{isobarplot}}
\end{figure}

The discontinuity of decay amplitude  crossing unitarity cut in a subenergy,  such as $s$, is given by 
\begin{align} 
\triangle  T(s,t,u) =& \frac{T(s+i \epsilon ,t, u ) - T(s-i \epsilon, t, u)}{2i} \nonumber \\
 =&     \rho(s) f^{*  }(s)   \frac{1}{2}  \int_{-1}^{1} d z_{s}     T (s, t, u), \label{fdis1}
\end{align}
where   \mbox{$\rho(s) = \sqrt{ 1- 4 m^{2}/s}$}, and $f(s)$ denotes  $S$-wave two-body elastic scattering amplitude and is parametrized by phase shift of two-body scattering, \mbox{$f= \left ( e^{2i\delta} - 1 \right ) /2i \rho  $}. \mbox{$z_{s} = \cos \theta_{s} $} is  given by  \mbox{$z_{s}= - (t - u  )/ \rho(s) k(s) $}, 
where \mbox{$k(s) = \sqrt{ \left [s - (M-m)^{2} \right ] \left [  s - (M+m)^{2}\right]}$}.
A diagrammatic representation of discontinuity relations in   Eq.(\ref{fdis1}) is shown in Fig.\ref{isobarplot}.

 \begin{figure}
\includegraphics[width=0.48\textwidth]{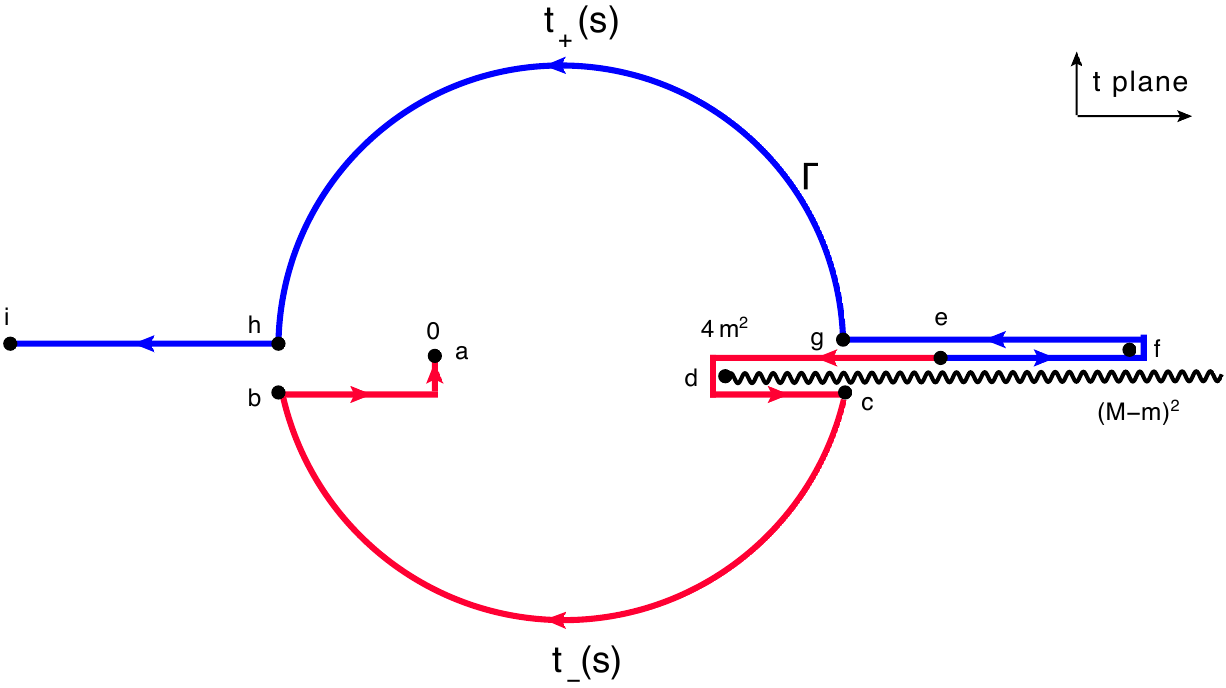}
\caption{   The path of $t_{\pm}(s)$  in $t$ complex plane as $s$ increased from $4 m^{2}$ to $\infty$.   The black waggle lines represent  right hand cuts of $g(t)$ function.  The  points labeled by \mbox{$a-i$} correspond to (a)\mbox{$t_{-}( \infty)= 0$}, (b)\mbox{$t_{-}\left( (M+m)^{2}\right)=  m (m-M)$},  (c)\mbox{$t_{-}\left((M-m)^{2} \right)=  m(M+m)  $}, (d)\mbox{$t_{-}\left( (M^{2} -m^{2})/2 \right)=4 m^{2}$}, (e)\mbox{$t_{\pm}( 4 m^{2})= ( M^{2} -m^{2} )/2$}, (f)\mbox{$t_{+}\left ( m(M+m) \right)= (M-m)^{2}$},  (g)\mbox{$t_{+}\left( (M-m)^{2} \right)=  m(M+m)$}, (h)\mbox{$t_{+}\left((M+m)^{2}\right)=  m(m-M)$},  and (i)\mbox{$t_{+}(\infty)= -\infty$}, respectively.
\label{fig:2}}
\end{figure}

 \begin{figure}
\includegraphics[width=0.48\textwidth]{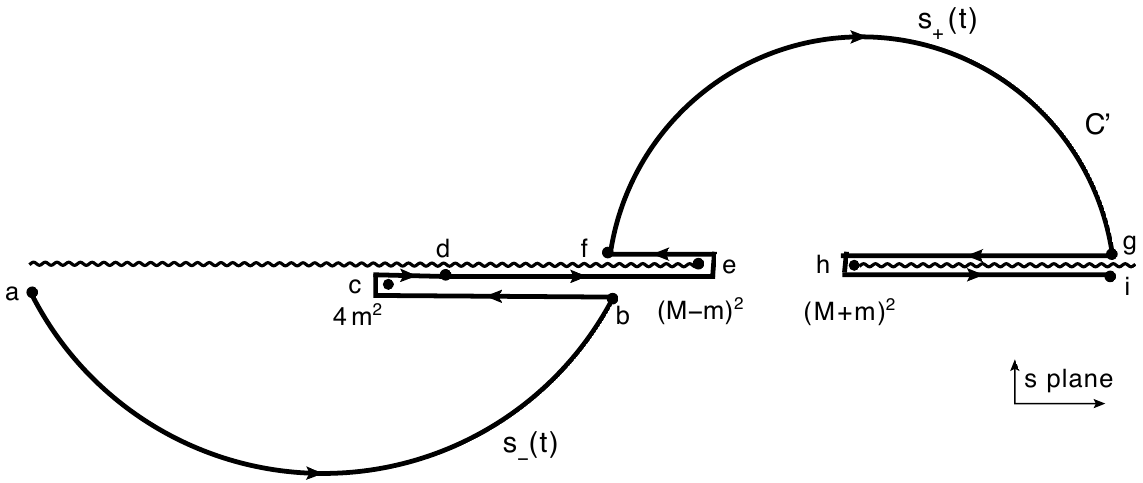}
\caption{   The path of $s_{\pm}(t)$    in $s$ complex plane for \mbox{$t \in \left [  -\infty, (M-m)^{2} \right ]$}.  The arrows indicate the directions  that   invariants follow along the path of integrations.
  The  black wiggle lines represent   cuts attached to two branch points: $(M\pm m)^{2}$ in $s$ plane  respectively. The  points labeled by \mbox{$a-f$} correspond to (a)\mbox{$s_{-}(0)= -\infty$}, (b)\mbox{$s_{-}(4 m^{2})= ( M^{2} -m^{2})/2$},  (c)\mbox{$s_{-}\left(( M^{2} -m^{2} )/2\right)= 4 m^{2}  $}, (d)\mbox{$s_{\pm}\left((M-m)^{2} \right)= m ( m+ M) $}, (e)\mbox{$s_{+}\left(m(M+m)\right)= (M-m)^{2} $}, (f)\mbox{$s_{+}(4m^{2})= (M^{2} -m^{2})/2$},  (g)\mbox{$s_{+}(0)= \infty$}, (h)\mbox{$s_{+}\left(m(m-M)\right)= (M+m)^{2}$},  and (i)$s_{+}(-\infty)= \infty$, respectively.
\label{fig:3}}
\end{figure}

By assumption, $a$'s possess only unitarity cuts, thus, \mbox{$\triangle  T(s,t,u)  =\triangle  a(s)$}, and
\begin{align} 
\triangle & a(s) =     \rho(s) f^{*  }(s)  \left [  a(s ) +    \frac{2}{\rho(s) k(s)} \int_{t_{-}(s)}^{t_{+}(s)}  \!\!\!\!\!\!\!\!\ \!\!\!\!\!\!  (\Gamma) \  d t  \,   a( t ) \right ], \label{fdisinvar}
\end{align}
where  the factor $2$ in front of contour integral  takes into account the  contribution  for $u$-channel. As discussed in  \cite{Aitchison:1965zz,Aitchison:1966kt,Pasquier:1968zz,Guo:2014vya}, the angular projection in Eq.(\ref{fdis1}) is replaced by a contour integration in complex plane according to perturbation theory \cite{Bronzan:1963xn,Bronzan:1964zz}, contour $\Gamma$ is given in Fig.\ref{fig:2}. The boundaries of Dalitz plot, $t_{\pm}(s)$, are   given by the solutions of  \mbox{$\phi(s,t_{\pm})=0$}, where \mbox{$ \phi(s,t)=s t u- m^{2}(M^{2}-m^{2})^{2}$},   the analytic continuation of $t_{\pm}(s)$ in $s$ is specified by   $\Gamma$, see  Fig.\ref{fig:2}. The scalar function $a$ then is determined by subenergy dispersion relation,
\begin{align}\label{disprelF}
a(s) &=  \frac{1}{\pi} \int_{4 m^{2}}^{\infty} d s' \frac{1}{s'-s   }\triangle  a(s')  .
\end{align}

As discussed in \cite{Guo:2014vya}, usually, it is useful to parameterize $a$ as a product of a known function and a reduced amplitude. 
For instance, we may choose parameterization,  \mbox{$a(s) = f(s) g(s)$}, thus, the discontinuity relation for the reduced amplitude $g$ is given by \cite{Guo:2014vya},
\begin{align}\label{discangg} 
  \triangle   g (s)    
= &-  \theta(s_{L}-s)    \frac{ \mbox{Im} f (s)  }{f^{*}(s)} g  (s)    \nonumber \\
 &+ \theta(s - 4 m^{2})     \frac{2}{ k(s)} \int_{t_{+}(s)}^{t_{-}(s)} \!\!\!\!\!\!\!\!\ \!\!\!\!\!\!  (\Gamma) \  d t  \,      f(t)  g(t) ,
\end{align}
where $s_{L}$ labels branch point of left hand cut in $f(s)$, and
\begin{align}\label{disprelg}
& g(s)   = \frac{1}{\pi} \int_{-\infty}^{\infty} d s'  \frac{1}{s'-s }   \triangle   g (s')  .
\end{align} 
By using Pasquier inversion technique  \cite{Aitchison:1965zz,Aitchison:1966kt,Pasquier:1968zz,Guo:2014vya}, also see Appendix~\ref{pasquier},  we may obtain a single integral equation for $g$,
 \begin{align}\label{pasqdisprelg}
 g(s)   = &- \frac{1}{\pi} \int_{-\infty}^{s_{L}} d s'  \frac{1}{s'-s}  \frac{ \mbox{Im} f (s')  }{f^{*}(s')}  g(s') + g_{R}(s)  ,
\end{align} 
where   
\begin{align} 
 g_{R}(s)  = \frac{2}{\pi} \int_{- \infty}^{(M-m)^{2}} \!\! d t  \, f(t)g(t)   \left [ \theta(t) \Delta(s, t ) -\theta(-t) \Sigma(s, t )  \right ]    . \label{physgR}
\end{align} 
  The kernel functions   $\Delta$ and $\Sigma$ are given by
     \begin{align}
\Delta ( s,t) &= \int_{s_{-}(t)}^{s_{+}(t)}  \!\!\!\!\!\!\!\!\ \!\!\!\!\!\!  (C') \ d s'  \frac{ 1 }{  U(s')  }  \frac{1}{s'-s } , \\
\Sigma  ( s,t) &= \int_{s_{+} (t)}^{ s_{+} (\infty)}  \!\!\!\!\!\!\!\!\!\!\!\!\!\!  (C') \  d s'  \frac{ 1 }{  U(s')  }  \frac{1}{s'-s }    ,
\end{align}
where   square root function   \mbox{$U(s)=\sqrt{ \left [s - (M-m)^{2} \right ]  \left [ s -(M+m)^{2} \right ]}$} is defined in complex-$s$ plane, the phase convention for $U(s)$ is chosen  by \mbox{$U(s\pm  i0 ) = (\mp , i , \pm)|U(s)|$} for \mbox{$s\in ([-\infty, (M-m)^{2} ]$}, \mbox{$[ (M-m)^{2} , (M+m)^{2} ], [ (M+m)^{2} , \infty])$} respectively.  Thus,  the square root $k(s)$   is given by  the value of $U(s)$   right below two cuts attached to \mbox{$(M\pm m)^{2}$},  \mbox{$k(s) =U(s-i0) $}. The contour $C'$ is given in Fig.\ref{fig:3}, and \mbox{$s_{\pm}(t)$} are specified by solutions of \mbox{$\phi(s_{\pm}, t)=0$} and contour $C'$.

The Pasquier inversion representation of $g(s)$ in Eq.(\ref{pasqdisprelg}-\ref{physgR}) is   initially defined in the range \mbox{$s \in [-\infty, (M-m)^{2}]$} (on left   and upper   side of contour $C'$).  As will be made clear in  section \ref{analyticpasquier}, contour $C'$ in kernel functions, $\Delta$ and $\Sigma$,  is singular and divides $s$ plane into several isolated regions.  Therefore, Eq.(\ref{pasqdisprelg}-\ref{physgR}) can only hold for a complex   $s$ that stays at the same side of contour $C'$ and does not cross contour $C'$. When $s$ is taken to cross contour  $C'$ to reach the region on the other side,  for Pasquier inversion representation of KT equation  to stay on physical sheet,  $C'$ has to be deformed and an extra piece  is picked up as the consequence of  deformation of contour. In follows, we present procedure of analytic continuation of Pasquier inversion representation of KT equation   into  \mbox{$s\in[(M-m)^{2}, \infty]$} regions.

 \begin{figure}
\includegraphics[width=0.45\textwidth]{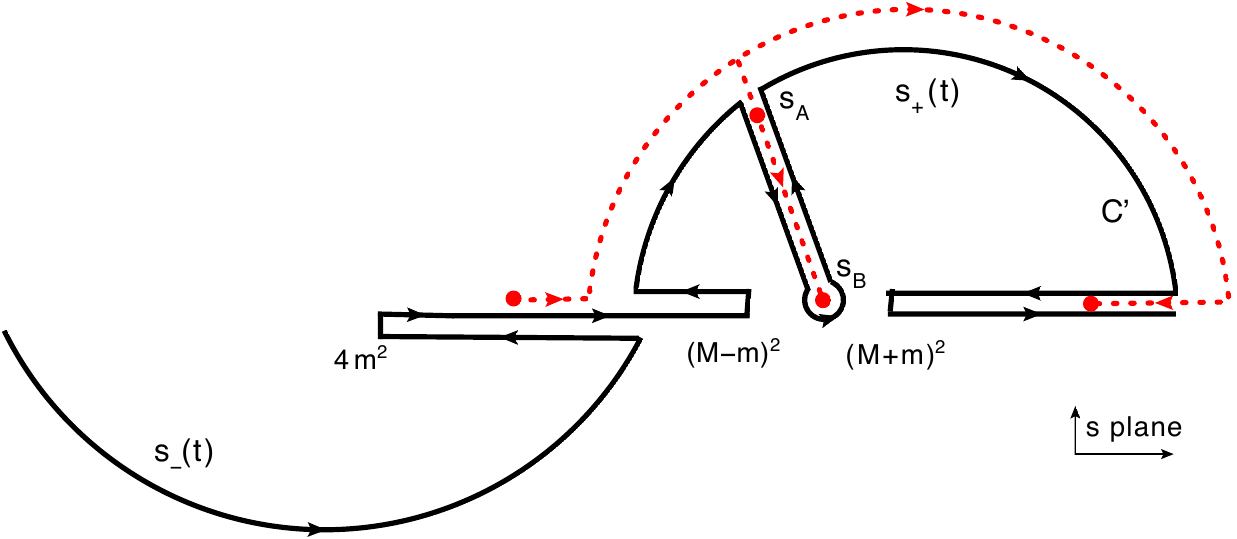}
\caption{   Analytic continuation of  a function of type,  $g_{R}(s)$ in Eq.(\ref{orderexchange}), is followed by the motion of $s$ (red dashed curve).   The physical sheet  of $g_{R}(s)$ is defined in the upper half plane that is divided by contour $C'$. Lower half  plane can be reached by crossing $C'$, when it does, a discontinuity   has to be picked up to keep $g_{R}(s)$  on physical sheet.
\label{pasqtriangle}}
\end{figure}

\section{Analytic continuation of Pasquier inversion representation for $s\in[(M-m)^{2}, \infty]$}\label{analyticpasquier}
As mentioned previously,   $g(s)$  given by  Eq.(\ref{pasqdisprelg}-\ref{physgR}) is originally     defined  for \mbox{$s\in[-\infty,(M-m)^{2}]$}. The analytic continuation of first term on right hand side of  Eq.(\ref{pasqdisprelg}) shows no  difficulty, therefore we will only focus on the second term on right hand side of  Eq.(\ref{pasqdisprelg}), $g_{R}(s)$, in following discussion. The $s$ dependence of $g(s)$ on  second term, $g_{R}(s)$, is through   kernel functions $\Delta $ and $\Sigma$,     $\Delta $ and $\Sigma$ on physical sheet  for \mbox{$s\in[- \infty,(M-m)^{2}]$} are   given by the value of $s$ running along the black wiggle line   attached to  $(M-m)^{2}$ in Fig.\ref{fig:3}. Therefore, the strategy of analytic continuation is that we start from here and then increase $s$ continuously until a singularity   is encountered.     Unfortunately for $g_{R}(s)$, contour $C'$   presents a cut in complex-$s$ plane which  stops us to naively use Eq.(\ref{pasqdisprelg}-\ref{physgR}) in nearby region \mbox{$s\in [(M-m)^{2},(M+m)^{2}]$}. To illustrate this point,  we first use  the   technique presented in Appendix~\ref{pasquier},  and rewrite   $g_{R}(s)  $  to,
 \begin{align} 
&g_{R}(s) = \frac{2}{\pi} \int_{\Gamma'}     d t \, a(t)   \int_{s_{\Gamma'}(t)}^{ \infty}  \!\!\!\!\!\!\!\!\!\!  (C') \ d s'   \frac{1}{s'-s }  \frac{ 1 }{  U(s')  }    , 
\end{align} 
where contour $\Gamma'$ is given in Fig.\ref{collapsepath}, the location of  $s_{\Gamma'}(t)$ on $C'$ is specified by  the location of $t$   on $\Gamma'$, see more details in Appendix~\ref{pasquier}. 
Exchanging the order of two integrals  leads to,
 \begin{align} 
& g_{R}(s) =\frac{2}{\pi} \int_{C'}   d s'   \frac{1}{s'-s }  \frac{ 1 }{  U(s')  }    \int_{0}^{ t_{C'}(s')}  \!\!\!\!\!\!\!\!\!\!\!\!\!  (\Gamma') \  \   d t \, a(t)   , \label{orderexchange}
\end{align} 
where $t_{C'}(s')$ is inverse of $s_{\Gamma'}(t)$, and Eq.(\ref{orderexchange}) is similar to Eq.(\ref{pathinteg}) but with   contours $C'$ and $\Gamma'$ instead.  Now, we can clearly see the cut structure on $s$ generated by contour $C'$ in Eq.(\ref{orderexchange}).  As $s$  is moved from left hand side of $C'$ in region  \mbox{$s\in[-\infty, (M-m)^{2}]$} to reach \mbox{$s\in[(M-m)^{2},(M+m)^{2}]$} region  by crossing   contour $C'$  (motion of $s$ is  demonstrated in Fig.\ref{pasqtriangle} by red dashed curve), $C'$ has to be deformed to keep $g_{R}(s)$ on physical sheet. For a example,   at a point $s_{A}$  in Fig.\ref{pasqtriangle}, which  sits right next to the inside circle of $C'$ in complex plane, then,  $g_{R}(s_{A})$ on physical sheet is given by,
 \begin{align} 
g_{R}(s_{A}) =& \frac{2}{\pi} \int_{C'}   d s'   \frac{1}{s'-s_{A} }  \frac{ 1 }{  U(s')  }    \int_{0}^{ t_{C'}(s')}  \!\!\!\!\!\!\!\!\!\!\!\!\!  (\Gamma') \  \   d t \, a(t)   \nonumber \\
+&   \frac{ 4 i }{  U(s_{A})  }    \int_{0}^{ t_{C'}(s_{A})}  \!\!\!\!\!\!\!\!\!\!\!\!\!\!\!  (\Gamma') \  \   d t \, a(t). \label{collapsegR}
\end{align} 
Next,   $s$ is moved away from $s_{A}$     to a point on real axis in region \mbox{$s \in [(M-m)^{2}, (M+m)^{2}]$}, such as \mbox{$s_{B}$} in Fig.\ref{pasqtriangle},  thus  $C'$ is further deformed to follow the motion of $s$. When $s$ reach real axis, $C'$ in second term on the right hand side of Eq.(\ref{collapsegR}) collapse onto   real axis and $\Gamma'$   opens up accordingly into $\Gamma$, thus, for \mbox{$s \in [(M-m)^{2}, (M+m)^{2}]$}  on the real axis, we obtain,
 \begin{align} 
g_{R}(s) =& \frac{2}{\pi} \int_{- \infty}^{(M-m)^{2}} \!\! d t  \, a(t)   \left [ \theta(t) \Delta(s, t ) -\theta(-t) \Sigma(s, t )  \right ]   \nonumber \\
+&   \frac{ 4 i }{  U(s)  }    \int_{0}^{ t_{+}(s)}  \!\!\!\!\!\!\!\!\!\!\!  (\Gamma) \   d t \, a(t), \nonumber \\
& \quad \quad \quad \ \ \mbox{for} \ \ s \in [(M-m)^{2}, (M+m)^{2}].
\end{align} 
At last, the analytic continuation of $g_{R}(s)$ from \mbox{$s\in[-\infty, (M-m)^{2}]$} to region \mbox{$s\in [(M+m)^{2}, \infty]$}, where $s$ runs along   black wiggle line attached to $(M+m)^{2}$ in Fig.\ref{fig:3}, does not encounter any singularities and so it does not  require the deformation of contour $C'$, see the motion of red dashed curve in  Fig.\ref{pasqtriangle}, therefore $g_{R}(s)$ in Eq.(\ref{physgR}) remains unchanged  for  \mbox{$s\in[ (M+m)^{2}, \infty]$}.

On the other hand, we may also perform the analytic continuation of Pasquier inversion representation of KT equation through a triangle diagram. Using Eq.(\ref{disprelF}),    we  first rewrite Eq.(\ref{pasqdisprelg}-\ref{physgR}) to 
 \begin{align}\label{cauchypasqdisprelg}
 g(s)   = &- \frac{1}{\pi} \int_{-\infty}^{s_{L}} d s'  \frac{1}{s'-s }  \frac{ \mbox{Im} f (s')  }{f^{*}(s')}  g(s') \nonumber \\
&+ \frac{2}{\pi} \int_{4 m^{2}}^{\infty}  d t' \triangle a(t') \mathcal{G} (s,t'),  \nonumber \\
& \quad \quad \quad \quad \quad \quad \quad \quad \mbox{for} \ \  s <(M-m)^{2},
\end{align} 
where  $\mathcal{G}  $  is given by,
  \begin{align}\label{trianglediagram}
 \mathcal{G}(s,t')   = &   \frac{1}{\pi} \int_{- \infty}^{(M-m)^{2}}  \!\! d t   \frac{1}{t'-t}   \left [ \theta(t) \Delta(s, t ) -\theta(-t) \Sigma(s, t)  \right ]    , \nonumber \\
&\quad \quad  \quad \quad      \mbox{for} \ \ t'>4m^{2}, s <(M-m)^{2}.
\end{align} 
$\mathcal{G} $  is identified as the Pasquier inversion  representation of a triangle diagram in region \mbox{$t'>4m^{2},s<(M-m)^{2}$}.   The analytic continuation of $\mathcal{G} $ in different representations     is presented in Appendix \ref{triangle}, the Pasquier inversion representation of $\mathcal{G} $  for \mbox{$(s,t')\in [-\infty,\infty]$} is  given by Eq.(\ref{fullpastrianglediagram}),
  \begin{align} 
 \mathcal{G}(s,t')    
 &=   \frac{1}{\pi} \int_{- \infty}^{(M-m)^{2}} d t  \frac{1}{t'-t}    \left [ \theta(t) \Delta(s, t) -\theta(-t) \Sigma(s, t)  \right ]    \nonumber \\
&+  2 i \theta  \left  (s - (M-m)^{2} \right ) \theta \left ((M+m)^{2} -s \right )     \nonumber \\
&  \quad   \times \left [ \frac{1 }{ U(s)} \int_{0}^{t_{+}(s)}  \!\!\!\!\!\!\!\!\! \!\! (\Gamma) \    \frac{ d t}{t'-t} +  \theta(t') \theta(4 m^{2} -t') \frac{2\pi i}{U(s)} \right ] , \nonumber \\
&\quad \quad  \quad \quad \quad \quad \quad \quad \quad \quad  \quad    \mbox{for} \ \ (s,t') \in [-\infty, \infty]. \nonumber
\end{align} 
The $s$  dependence of $g(s)$ in second term in Eq.(\ref{cauchypasqdisprelg}) is all through triangle diagram $\mathcal{G}  $, thus, analytic continuation of  $\mathcal{G}  $ completes the analytic continuation of Pasquier inversion representation of $g(s)$. Plugging Eq.(\ref{fullpastrianglediagram}) back into Eq.({\ref{cauchypasqdisprelg}}), we once again obtain the  Pasquier inversion representation of $g(s)$ for \mbox{$s\in [-\infty,\infty]$},
\begin{align}\label{analyticpasqdisprelg}
 g(s)   &= - \frac{1}{\pi} \int_{-\infty}^{s_{L}} d s'  \frac{1}{s'-s }  \frac{ \mbox{Im} f (s')  }{f^{*}(s')}  g(s') \nonumber \\
&+ \frac{2}{\pi} \int_{- \infty}^{(M-m)^{2}}  \!\! d t   f( t )  g(t)    \left [ \theta(t) \Delta(s, t ) -\theta(-t) \Sigma(s, t )  \right ]  \nonumber \\
&+ 4 i  \theta  \left  (s- (M-m)^{2} \right ) \theta \left ((M+m)^{2} -s \right )   \nonumber \\
 & \quad \quad   \quad \quad  \quad \quad  \quad \quad  \quad \times  \frac{1 }{ U(s)} \int_{0}^{t_{+}(s)}  \!\!\!\!\!\!\!\!\! \!\! (\Gamma)  \   d t  f( t )  g(t) ,  \nonumber  \\
&\quad \quad  \quad \quad \quad \quad \quad \quad \quad     \mbox{for} \ \ s \in [-\infty, \infty].
\end{align}

 \begin{figure}
\includegraphics[width=0.54\textwidth]{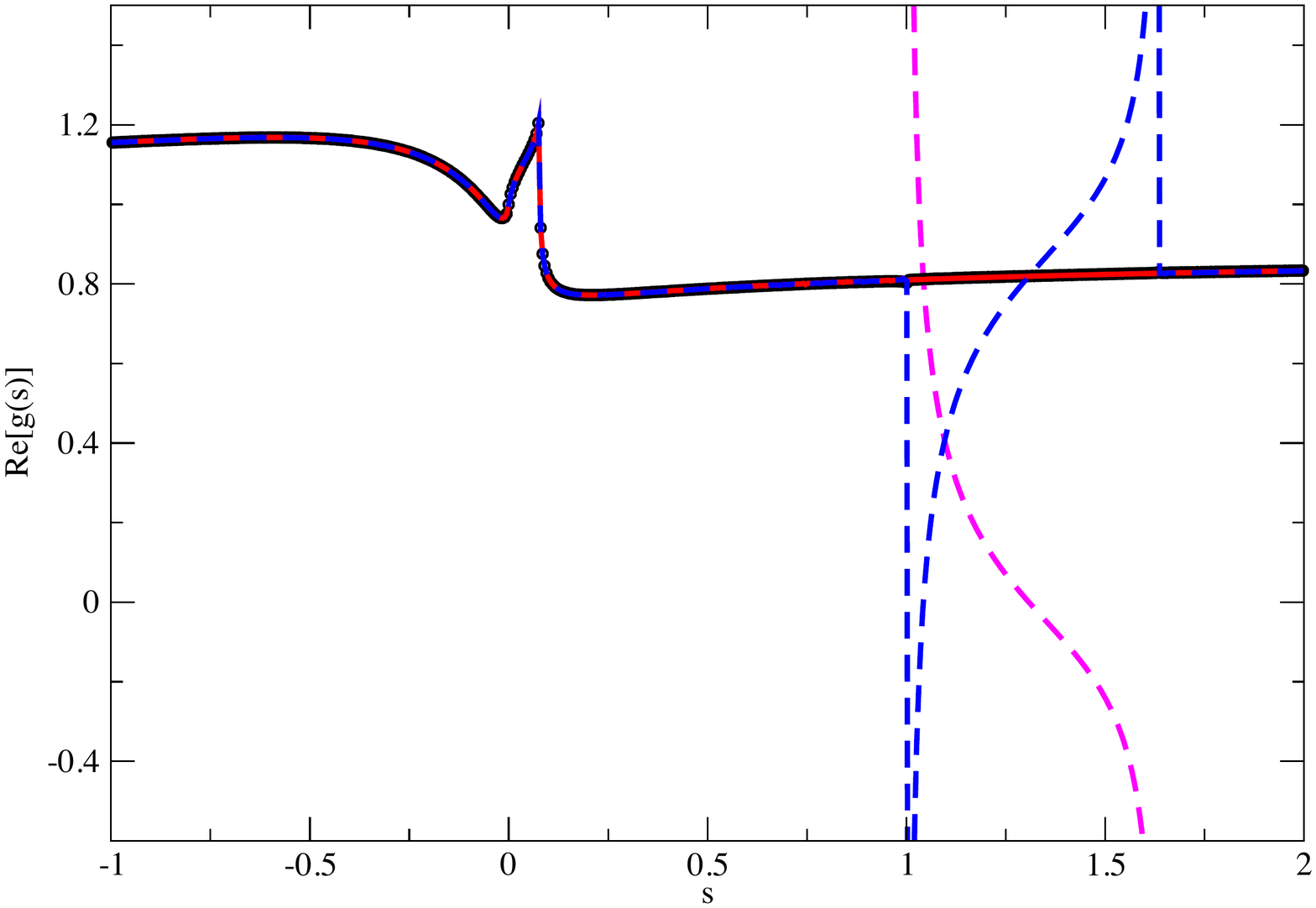}
\includegraphics[width=0.54\textwidth]{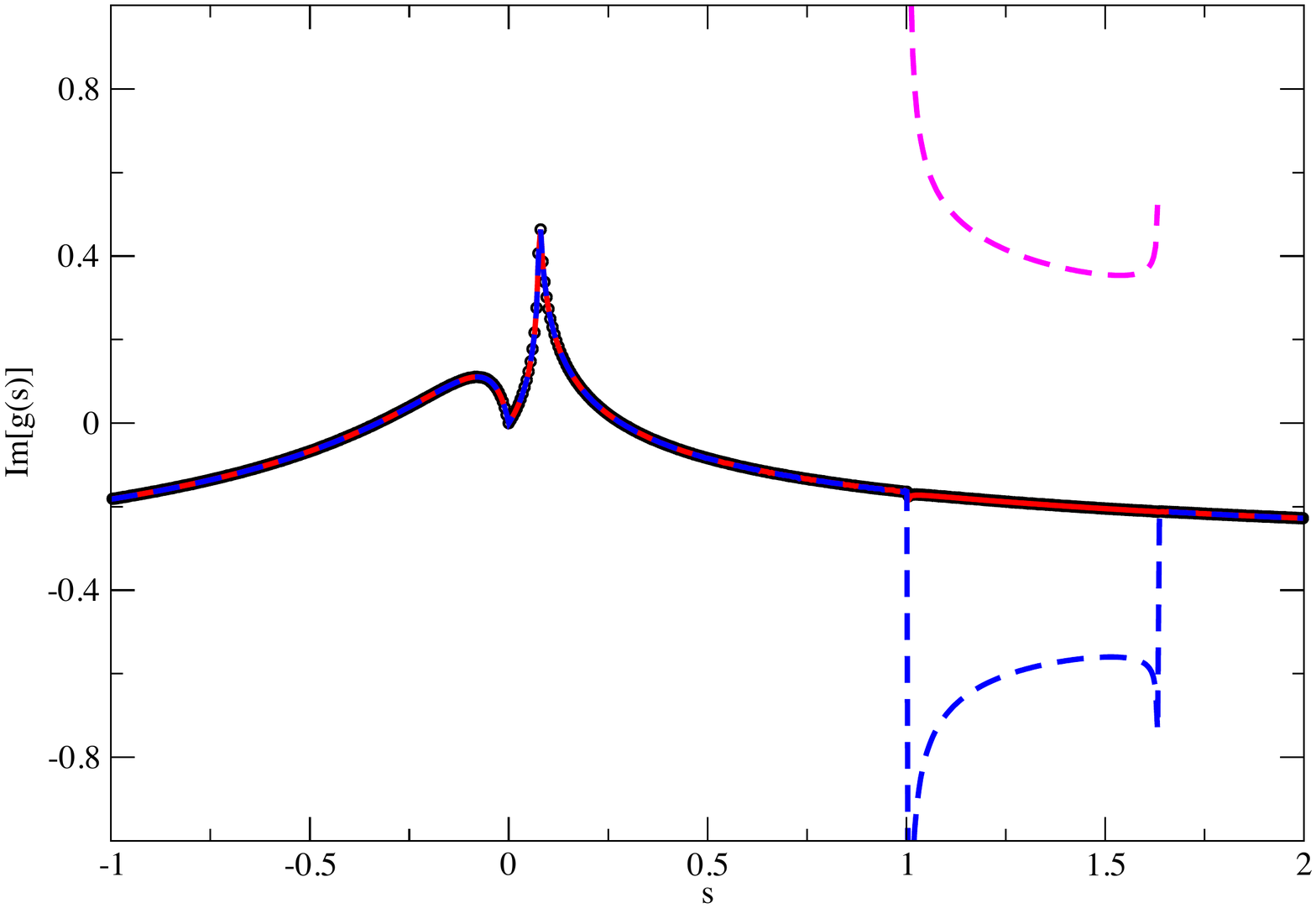}
\caption{  The real (top plot) and imaginary (bottom plot) parts of $g(s)$ by solving both dispersion representation Eq.(\ref{discangg}-\ref{disprelg}) (black circles) and Pasquier inversion representation Eq.(\ref{analyticpasqdisprelg}) (solid red curves). Blue dashed curves are the solution of $g(s)$ from Eq.(\ref{pasqdisprelg}-\ref{physgR}) without proper analytic continuation, purple dashed  curves represent contribution from extra term picked up by analytic continuation in Eq.(\ref{analyticpasqdisprelg}),  $4 i /U(s) \int_{0}^{t_{+}(s)}   (\Gamma)    d t  a(t) $. The input model of $f(s)$ is taken from Eq.(28) in \cite{Guo:2014vya}, with fixed parameters: \mbox{$\alpha=0.1$}, \mbox{$\beta=0.2$}, \mbox{$m_{R}=0.8$ GeV}, \mbox{$m=0.14$ GeV} and \mbox{$M=1.14$ GeV}. The left hand cut of $f(s)$ is placed at \mbox{$s_{L}=0$} and $g(s)$ is normalized to \mbox{$g(0)=1$}.
\label{fig:gcont}}
\end{figure}

The analytic continuation of Pasquier inversion representation of $g(s)$ given by  Eq.(\ref{analyticpasqdisprelg}) is tested   numerically by using a toy model for $f(s)$,     model of $f(s)$ is taken from \cite{Guo:2014vya}.  The comparison of $g(s)$'s by solving  Pasquier inversion representation Eq.(\ref{analyticpasqdisprelg})  and dispersion representation Eq.(\ref{discangg}-\ref{disprelg}) is shown in Fig.\ref{fig:gcont}. We also shown the results by solving Eq.(\ref{pasqdisprelg}-\ref{physgR}) without proper analytic continuation compared to the contribution of  extra term that is picked up by analytic continuation, \mbox{$4 i/U(s)\int_{0}^{t_{+}(s)}   (\Gamma)    d t  a(t) $}. As demonstrated in Fig.\ref{fig:gcont},   solution of analytic continuation of Pasquier inversion representation of $g(s)$   is consistent with  dispersion representation of $g(s)$.  Solution of Pasquier inversion representation of $g(s)$ without proper analytic continuation jumps  in unphysical region \mbox{$s\in[(M-m)^{2},(M+m)^{2}]$},  extra term, \mbox{$4 i /U(s) \int_{0}^{t_{+}(s)}   (\Gamma)    d t  a(t) $}, is needed to keep $g(s)$ continuous and staying on physical sheet.

At last, similarly,  if we parametrize \mbox{$a(s) = G(s)/D(s)$} \cite{Guo:2014vya}, where $D(s)=N(s)/f(s)$ contains only unitarity cut of scattering amplitude and all other cuts are absorbed into function $N(s)$ \cite{Chew:1960iv,Frye:1963zz}, thus we obtain,
 \begin{align} 
 G(s)   =& \frac{2}{\pi} \int_{- \infty}^{(M-m)^{2}} \!\!  d t     \frac{G(t)}{D(t)}    \left [ \theta(t) \Delta_{G}(s, t ) -\theta(-t) \Sigma_{G}(s, t )  \right ]  \nonumber \\
+& 4 i  \theta  \left  (s- (M-m)^{2} \right ) \theta \left ((M+m)^{2} -s \right )   \nonumber \\
 &  \quad \quad   \quad \quad  \quad \quad  \quad \quad   \times  \frac{N(s) }{ U(s)} \int_{0}^{t_{+}(s)}  \!\!\!\!\!\!\!\!\! \!\! (\Gamma)  \   d t \frac{G(t)}{D(t)}    , \nonumber  \\
&\quad \quad  \quad \quad \quad \quad \quad \quad       \mbox{for} \ \ s \in [-\infty, \infty],  \label{analycontdispersionrep}
\end{align} 
where the kernel functions $\Delta_{G}$ and $ \Sigma_{G}$ are given by Eq.(\ref{deltaG}) and (\ref{deltaG}) respectively.

\section{Summary} \label{summary}
We presented the analytic continuation procedure of Pasquier inversion representation of KT equation, and a well-defined Pasquier inversion representation of KT equation for an arbitrary $s$ on real axis  is given by Eq.(\ref{analyticpasqdisprelg}) and Eq.(\ref{analycontdispersionrep}).     

Comparing the Pasquier inversion representation of KT equation in Eq.(\ref{analyticpasqdisprelg}) to dispersion representation of KT equation in Eq.(\ref{discangg}-\ref{disprelg}), as has been also discussed in \cite{Guo:2014vya}, the single integral form of Pasquier inversion    representation in Eq.(\ref{analyticpasqdisprelg})   indeed present a significant   advantage on numerical computation in regions \mbox{$s\in[-\infty,(M-m)^{2}]$} and \mbox{$s\in[(M-m)^{2},\infty]$}. However, in unphysical region \mbox{$s\in[(M-m)^{2},(M+m)^{2}]$},  dispersion representation in Eq.(\ref{discangg}-\ref{disprelg}) requires no extra efforts, but analytic continuation of Pasquier inversion representation becomes non-trivial and need an extra term to keep solution, $g(s)$, staying on physical sheet. At last, we solved  Pasquier inversion representation of  KT equation in Eq.(\ref{analyticpasqdisprelg}) numerically by using a toy model of $f(s)$,      solutions  with and without proper analytic continuation compared to the solution of dispersion representation  are illustrated in Fig.\ref{fig:gcont}.

\section{ACKNOWLEDGMENTS}
We thank Adam P. Szczepaniak for  many fruitful discussion. We acknowledge supports in part   by the U.S.\ Department of Energy under Grant No.~DE-FG0287ER40365 and the Indiana University Collaborative Research Grant. We also   acknowledge support from U.S. Department of Energy contract DE-AC05-06OR23177, under which Jefferson Science Associates, LLC, manages and operates Jefferson Laboratory.

\appendix

\section{Pasquier inversion technique}\label{pasquier}

 \begin{figure}
\includegraphics[width=0.48\textwidth]{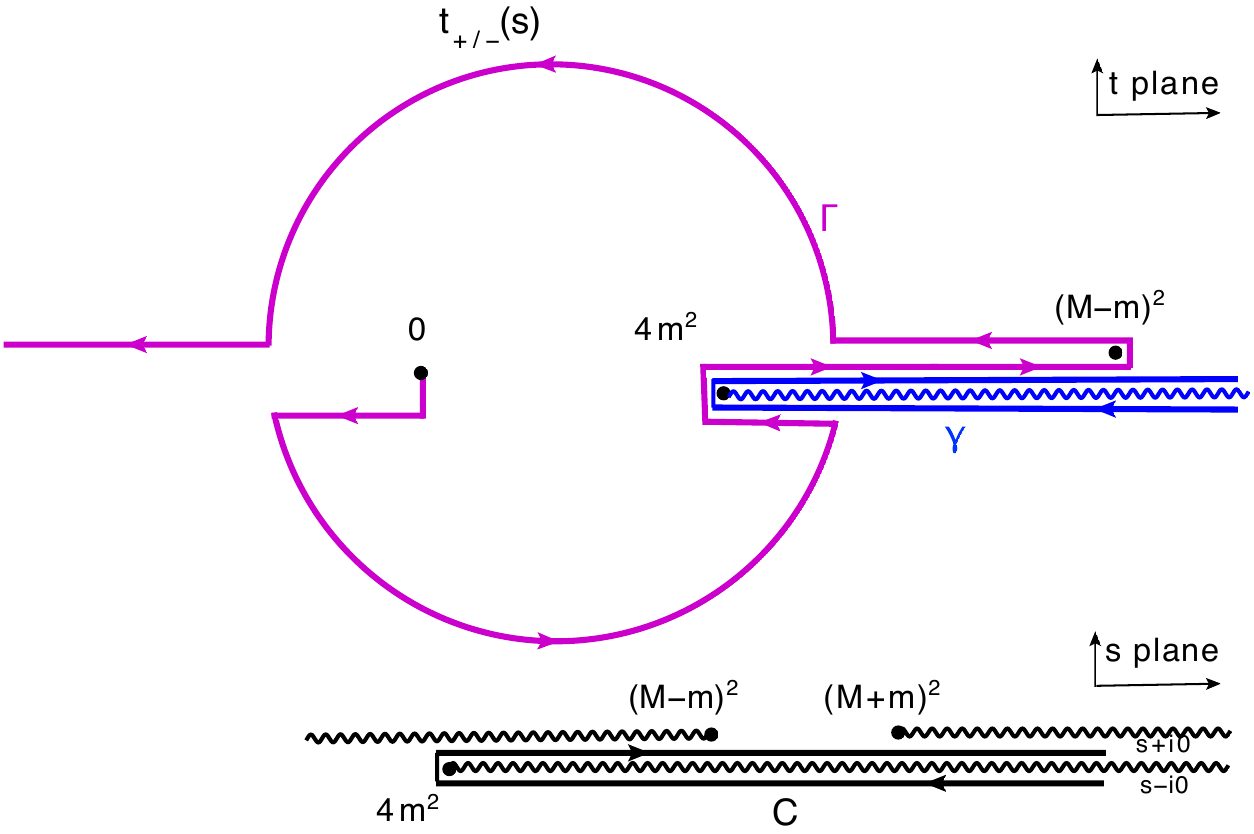}
\caption{   The path of $t_{\pm}(s)$ (solid purple  curves),      contour $C$ (solid black  lines) and $\gamma$ (solid blue  lines).  The arrows indicate the directions  that   invariants follow in double integrations, Eq.(\ref{doubleintef}) and (\ref{inverseintef}). The blue wiggle line represents the unitarity cut in $t$ plane, and black wiggle lines represent  the cuts in $s$ plane.
\label{doublepath}}
\end{figure}

 \begin{figure}
\includegraphics[width=0.48\textwidth]{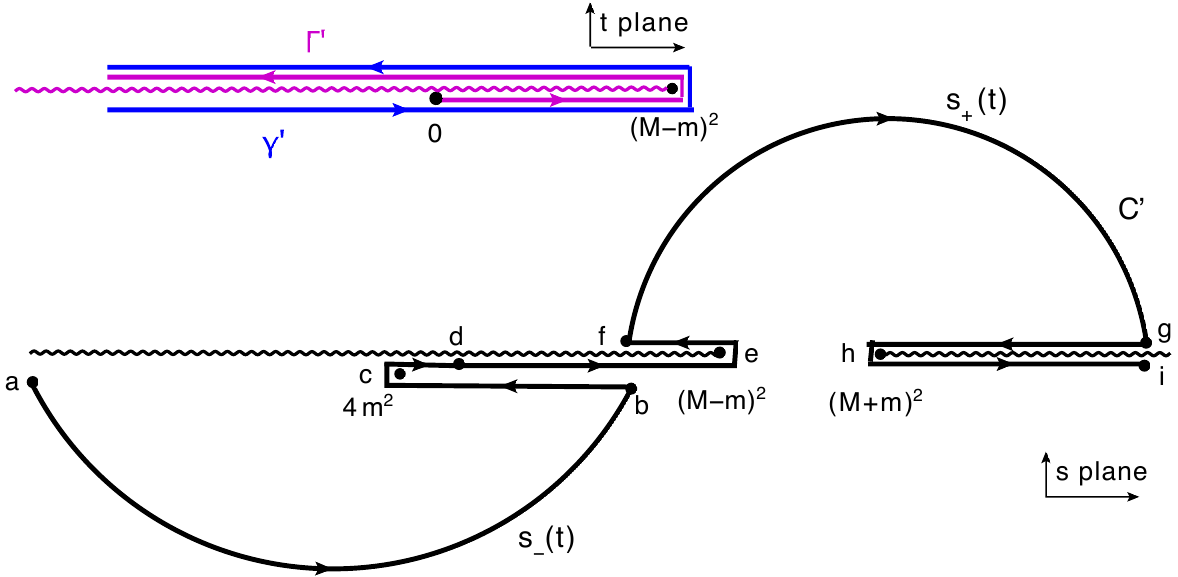}
\caption{   The contour $\Gamma'$ (solid purple lines),       $C'$ (solid black curves) and $\gamma'$ (solid blue lines) in Eq.(\ref{collapseintef}).  The arrows indicate the directions  that   invariants follow along the path of integrations.  The purple and black wiggle lines represent unitarity cut in $s$ plane and  cuts attached to two branch points: $(M\pm m)^{2}$ in $s$ plane  respectively. 
\label{collapsepath}}
\end{figure}

 For completeness,  we present the Pasquier inversion technique \cite{Pasquier:1968zz,Aitchison:1978pw} in this section. Considering a   double integrals equation of   type, 
 \begin{equation}\label{doubleintef}
I(s)=  \int_{ 4 m^{2}}^{\infty} d s' \frac{1}{s'-s} \frac{N(s') }{k(s')}   \int_{t_{-}(s')}^{t_{+}(s')}  \!\!\!\!\!\!\!\!\!\!\! (\Gamma) \    d t \, a(t),
\end{equation}
where contour $\Gamma$  followed by $t$ integration  is defined to avoid unitarity cut in $a$, see  Fig.\ref{fig:2}, and the integration path of $s'$ is defined on the real axis,      the physical value of $I(s)$   is given by $s$  running  above real axis.

As described in  \cite{Pasquier:1968zz,Aitchison:1978pw}, we first split $ t$ integral into two pieces and rewrite  the double integrals in Eq.(\ref{doubleintef}) to 
\begin{equation}\label{splitintef}
I(s)=   \int_{4 m^{2}}^{\infty} d s'  \frac{1}{s'-s} \frac{N(s')  }{k(s')} \left [ \int_{0}^{t_{+}(s')}  \!\!\!\!\!\!\!\!\!\!\!\!  (\Gamma)    \  -   \int_{0}^{t_{-}(s')}  \!\!\!\!\!\!\!\!\!\!\!\! (\Gamma)   \ \  \right ]  d t \, a(t).
\end{equation}
Then, for first term in bracket in Eq.(\ref{splitintef}),  the path of $s'$ integration is shifted to above   real axis, and for second term in  bracket in Eq.(\ref{splitintef}),  the path of $s'$ integration is shifted to below   real axis.
 Note that  kinematic function $k(s')$ as a function of $s'$   has two branch points: $(M \pm m)^{2}$.   Two cuts may be attached to  these two   points, one   runs  from $-\infty$ up to $(M -m)^{2}$ and another    runs from $(M + m)^{2}$ up to $\infty$, see black wiggle lines in Fig.\ref{doublepath}.  As we have mentioned before, $U$ is defined as the continuation of $k(s')$ for a complex argument, the physical value of $k(s')$   is given by taking the branch of $U(s')$     below two cuts attached to \mbox{$(M\pm m)^{2}$},  \mbox{$k(s') =U(s'-i0) $}. These two kinematic cuts are placed above both    real axis and     shifted     $s'$ integration paths  described previously, see Fig.\ref{doublepath}, therefore,   operation of shifting   $s'$ integration paths is valid and $s'$ integration paths  do not interfere with   cuts in $U$.      
Thus, we can safely   rewrite  the double integrals in Eq.(\ref{splitintef}) to, 
\begin{equation}\label{pathinteg} 
 I(s)= \int_{C} d s'   \frac{1}{s'-s} \frac{N(s')  }{U(s')}  \int_{0}^{t_{C}(s')}   \!\!\!\!\!\!\!\!\!\!\!\! (\Gamma)   \ \ d t \,  a(t),
\end{equation}
where the path $C$ of $s'$ integral is shown in Fig.\ref{doublepath}, whether $t_{C}(s')$ is \mbox{$t_{+}(s')$ or $t_{-}(s')$} depends on  which portion of path  $C$  the invariant $s'$ is on. $t_{+}(t_{-})$ is assigned  to  $s'$ on the portion of $C$ above(below) real axis.   In Eq.(\ref{pathinteg}), the physical value of $I(s)$ is given by $s$ running above contour $C$. 
 Next, we exchange the order of two integrals, so that Eq.(\ref{pathinteg}) becomes,
\begin{equation}\label{inverseintef} 
I(s)=   \int_{\Gamma} d t   \, a(t) \left [ \int_{s_{\Gamma} (t)}^{\infty}  \!\!\!\!\!\!\!\!\! (C)  \   d s'   \frac{1}{s'-s} \frac{N(s') }{U(s')} \right ],
\end{equation}
where $t$ integration   on contour $\Gamma$ runs from $0$ to $- \infty$ by looping around threshold $(M-m)^{2}$, see Fig.\ref{doublepath}, and $s'$ integration   runs from $s_{\Gamma}(t)$ up to $\infty$ along path  $C$, $s_{\Gamma}(t)$ is given by the inverse of $t_{C}(s')$.  By   assumption, $a $ has only unitarity cut,  using Cauchy's theorem, we can write a equation, 
\begin{equation}
a(t) = \frac{1}{2 \pi i} \int_{\gamma} d t' \frac{a(t')}{t'-t},
\end{equation}
where contour $\gamma$   loops around the unitarity cut but avoiding interference with $\Gamma$, see in Fig.\ref{doublepath},  the  convergence of integration has been assumed valid  so that the   circle of contour $\gamma$ at infinity can be dropped. Thus, we obtain,
\begin{align}\label{cauchyintef} 
   I(s)=&\frac{1}{2 \pi i}  \int_{\gamma} d t' \, a(t')   \left [  \int_{\Gamma} \frac{d t }{t'-t}   \int_{s_{\Gamma} (t)}^{\infty}  \!\!\!\!\!\!\!\!  (C) \  d s'     \frac{1 }{s'-s}  \frac{N(s') }{U(s')} \right ].
\end{align}
When $N$ function is replaced by a constant, the function in bracket in Eq.(\ref{cauchyintef})  may be associated to a  triangle diagram, see Appendix~\ref{triangle}.
 The next step is to deform the contour $\gamma$ onto   real axis toward \mbox{$-\infty$} but avoid both unitarity cut in $a$ and the singularities from the expression in bracket in Eq.(\ref{cauchyintef}).
 By construction of $\Gamma$, unitarity cut of $a$ sits along the blue wiggle line in Fig.\ref{doublepath},   and $a(t)$ for $t$ running above unitarity cut is defined  physical. Therefore, as long as   deformation of $\gamma$ and $\Gamma$ toward negative real axis does not interfere with unitarity cut of $a$,   $a$ remains on physical sheet all the time.
  Singularities of the function in bracket  in Eq.(\ref{cauchyintef}) have been extensively studied by authors in  \cite{Aitchison:1965zz,Pasquier:1968zz,Aitchison:1964ta,Aitchison:1964nc,Kacser:1966ta} from perturbation theory perspective,     they are branch points at \mbox{$t= 0,   4 m^{2}$} and $(M - m)^{2}$. One may attach  branch cuts to those branch points running toward negative real axis \cite{Aitchison:1965zz,Pasquier:1968zz,Aitchison:1964ta,Aitchison:1964nc,Kacser:1966ta},  therefore,   the   contour $\gamma'$   may be chosen   to loop around the threshold $(M - m)^{2}$   toward negative real axis. The deformation of contour $\gamma$ also drags the contour $\Gamma$ going with it back onto   real axis,    the correspondent  contour $C$ must then be opened up accordingly. Simultaneously, in order for $I(s)$ staying on physical sheet, some    $s$ are also dragged by the deformation of $C$ into complex plane, and physical value of $I(s)$ is now  given by a $s$ that sits on the same side of      $C$ when it opens up into complex plane.  The only  $t'$-dependent  singularities come from factor $1/(t'-t)$  in bracket in Eq.(\ref{cauchyintef}), so that, when contour $\gamma$ is collapsed onto real axis,   the discontinuity of  this factor $1/(t'-t)$  along the cut  from $-\infty$ to $(M- m)^{2}$ is picked up. Equivalently,  we may replace $  \int_{\gamma}  d t'  /(t'-t)$ by \mbox{$ 2\pi i  \int_{-\infty}^{ (M - m)^{2}} d t'  \delta( t'-t)$} in Eq.(\ref{cauchyintef}). Therefore, Eq.(\ref{cauchyintef}) becomes,
\begin{align}\label{collapseintef} 
 I(s)=&   \int_{  \Gamma'}  d t  \, a(t)     \left [  \int_{s_{\Gamma'} (t)}^{\infty}  \!\!  \!\!\!\!\!\!\!\! (C') \ d s'     \frac{1 }{s'-s} \frac{N(s') }{U(s')} \right ] .
\end{align}
The contours   $C'$  and $\Gamma'$ have to be chosen  to avoid the singularities in integrands.  Examining Eq.(\ref{collapseintef}),  we note that    integrands  of contour integration over invariant $s'$ are only the product of $1/(s'-s)$, kinematic function $U$ and left hand cut function $N$. 
As we have mentioned early, kinematic function $U$ as a function of $s'$  has two branch points: $(M \pm m)^{2}$,  two cuts are attached  to these two   points, one   runs toward $-\infty$ and another runs toward $\infty$ respectively, see black wiggle lines in Fig.\ref{collapsepath}.
 Therefore, the contour $C'$ may be chosen to avoid   two cuts attached to $(M\pm m)^{2}$, as plotted in  Fig.\ref{collapsepath}. With this choice, the function in bracket in Eq.(\ref{collapseintef}) may be associated to the discontinuities of triangle diagram defined  in Eq.(\ref{cauchyintef}) along the cut attached to branch point $(M- m)^{2}$ in $t$ plane.    Similar to $\gamma'$,  the contour  $\Gamma'$ also   loops around the   point    $(M- m)^{2}$ and is placed above unitarity cut in $a$.
Whether $s_{\Gamma'} (t)  $ is $s_{+} (t) $ or $s_{-} (t) $ depends on whether $t$ is above or below the cut attached to $(M-m)^{2}$ in $t$ plane respectively, see Fig.\ref{collapsepath}.  

As   mentioned early, the physical value of $I(s) $ was chosen by $s$ running above   contour $C$ in Eq.(\ref{pathinteg}), when $C$ is deformed to $C'$, see Fig.\ref{doublepath} and Fig.\ref{collapsepath}. In order  for $I(s)$ to stay on physical sheet, $s$ is not allowed to cross contour, as a result, some   $s$ are forced  to follow   the deformation of contour into complex plane.  Specifically, (1) for \mbox{$s \in [-\infty, (M-m)^{2}]$},   the physical value of $I(s)$ is given by $s$ running along the black wiggle line attached to \mbox{$(M-m)^{2}$}, (2) for \mbox{$s \in [ (M-m)^{2}, \infty]$}, now physical  $I(s)$ is trapped into the value of $s$ running along   balck wiggle line attached to \mbox{$(M+m)^{2}$} between sections \mbox{$g-h$} and \mbox{$h-i$} on $C'$, and (3)   the form of $I(s)$  in Eq.(\ref{collapseintef}) for    \mbox{$s\in  [(M-m)^{2}, (M+m)^{2}]$} on real axis is no longer on the physical sheet,   physical $I(s)$  now is given by a complex $s$ running on the upper side of arc \mbox{$f-g$} on $C'$. To reach physical $I(s)$ for \mbox{$s\in  [(M-m)^{2}, (M+m)^{2}]$} on real axis, the analytic continuation is required, the procedure is described in section \ref{analyticpasquier}.

At last,  by splitting   $s'$ integration path,  \mbox{$   \int_{ \Gamma'}   =  \left [  \int_{ 0 _{-}}^{ (M - m)^{2}_{-} }  -    \int^{ (M - m)^{2}_{+}}_{  0_{+} }  \right ]  +\int_{ 0_{+}}^{  - \infty_{+}}        $} in Eq.(\ref{collapseintef}) (subscript $+/-$ of integration limits denotes the path of integration lying above or below the cut attached to branch point $(M-m)^{2}$ in $t$ plane, see Fig.\ref{collapsepath}), we obtain,
\begin{align}\label{pasqinvert} 
  & I(s)= \int_{ -\infty }^{ (M - m)^{2} } \!\!\!\!   d t \,  a(t)      \left [   \theta(t)  \Delta_{G} (s, t) - \theta(-t)  \Sigma_{G} (s, t)   \right ]   ,
\end{align} 
  where  
    \begin{align}
\Delta_{G} ( s,t) &= \int_{s_{-}(t)}^{s_{+}(t)}  \!\!\!\!\!\!\!\!\ \!\!\!\!\!\!  (C') \ d s'   \frac{1}{s'-s } \frac{ N(s')  }{  U(s')  }  \label{deltaG}, \\
\Sigma_{G}  ( s,t) &= \int_{s_{+} (t)}^{ \infty}  \!\!\!\!\!\!\!\!\ \!\!\!  (C') \  d s'  \frac{1}{s'-s }  \frac{ N(s')  }{  U(s')  }   \label{sigmaG} .
\end{align}
For    $s$ on real axis,    value of $\Delta_{G} $ and $\Sigma_{G} $ on physical sheet is only defined  in regions, \mbox{$[-\infty,(M-m)^{2}]$} and \mbox{$[(M+m)^{2},\infty]$}. For \mbox{$s\in[(M-m)^{2}, (M+m)^{2}]$},   $\Delta_{G} $ and $\Sigma_{G} $ given by Eq.(\ref{deltaG}-\ref{sigmaG}) without proper analytic continuation   are   on unphysical sheet.  For the case \mbox{$N(s)=1$}, the corresponding kernels are denoted as $\Delta $ and $\Sigma $.

Kernel functions   $\Delta   $ and $\Sigma $   can be expressed in terms of elementary functions. For real $s$ and $t$, the value of   $\Delta   $ and $\Sigma $ given below by Eq.(\ref{deltaphy}-\ref{sigmaunphy}) are simply corresponding to the limit \mbox{$s+i0$} and \mbox{$t+i0$}, and again, Eq.(\ref{deltaunphy}) and (\ref{sigmaunphy}) are defined on unphysical sheet,
   \begin{align} 
 \Delta   (s, t) &  = \frac{1}{U(s )}   \ln | \frac{R  (s, t) + U(s ) U(t  )}{R  (s, t) - U(s   ) U(t  ) } |   - \theta \left ( \phi  (s, t) \right ) \frac{ i \pi}{U(s )} , \nonumber \\
 &       \mbox{for} \ \  t\in[0, (M-m)^{2}],  \nonumber \\
 &   \quad  \   \ s\in [-\infty, (M-m)^{2}] \& [(M+m)^{2}, \infty],   \label{deltaphy}
\end{align}
and
 \begin{align} 
 & \Delta   (s, t)  = \frac{1}{U(s )}   \ln  \frac{R  (s, t) + U(s ) U(t  )}{R  (s, t) - U(s   ) U(t  ) }    \nonumber \\
 & \quad \quad \quad \quad -   \theta\left (m(m+M)-t \right ) \theta \left ( s_{R}(t)-s  \right )  \frac{ 2 i \pi}{U(s )} , \nonumber \\
  &    \mbox{for}    \ \  t\in[0, (M-m)^{2}],   s\in [  (M-m)^{2}, (M+m)^{2}], \label{deltaunphy}
\end{align}
where \mbox{$R (s, t)  = - M^{4}  +  (s - m^{2}) (t- m^{2})    +M^{2} (s+t  )$} and $s_{R}(t)$ is given by the solution of \mbox{$R(s_{R},t)=0$}.
 \begin{align} 
&\Sigma (s, t) = \frac{1}{U(s )}    \ln | L(s,t)  |-    \theta \left(  s- s_{+}(t) \right)  \frac{  i \pi}{U(s )}  , \nonumber \\
 &           \mbox{for} \ \  t <0,  s\in  [-\infty, (M-m)^{2}] \& [(M+m)^{2}, \infty],    \label{sigmaphy}
\end{align} 
and
 \begin{align} 
& \Sigma (s, t) = \frac{1}{U(s )}    \ln  L(s,t)  -    \theta \left( \mbox{Im}L (s, t) \right)  \frac{ 2 i \pi}{U(s )}  , \nonumber \\
 &   \quad \quad  \quad               \mbox{for} \ \  t <0,  s\in [  (M-m)^{2}, (M+m)^{2}],   \label{sigmaunphy}
\end{align} 
 where 
\begin{align}
&L (s, t)  \nonumber \\
& = \frac{ \left [ s_{+}(t )-s  \right ] \left [ s- M^{2} -m^{2} + U(s) \right ]  }{ \left [ s_{+}(t )-s  \right ]  (s- M^{2} -m^{2} )  + U^{2} (s) - U(s) U ( s_{+} (t) )} . \nonumber
\end{align}

 \begin{figure}
\includegraphics[width=0.24\textwidth]{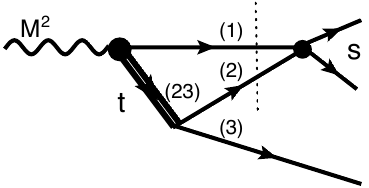}
\caption{   A triangle diagram with a fixed internal mass $\sqrt{t}$ in $(23)$ sub-channel.
\label{trianglediagram}}
\end{figure}

\section{Different representations of a triangle diagram}\label{triangle}
From perturbation theory, the Feynman parametrization of a triangle diagram in Fig.\ref{trianglediagram}   is given by \cite{Bronzan:1963xn},
\begin{align}\label{feynmantriangle} 
&  \mathcal{G}(s,t)  = \frac{1}{\pi} \int_{0}^{1} d \alpha_{1} d \alpha_{2} d \alpha_{3}  \nonumber \\
&\times \frac{\delta(1-\alpha_{1} - \alpha_{2} - \alpha_{3})}{\alpha_{1} t + (1- \alpha_{1} -\alpha_{1} \alpha_{2}) m^{2} - \alpha_{3} (\alpha_{1} M^{2} + \alpha_{2} s) -i \epsilon},
\end{align} 
where $t$ denotes the invariant mass square of  pair $(23)$ propagator. The analytic continuation of   Feynman parametrization representation of $\mathcal{G}$  as a function of complex arguments $(s,t)$ is carried out by $i \epsilon$ prescription \cite{Bronzan:1963xn}.

In follows, we   present the analytic continuation of both dispersion representation and Pasquier inversion representation of $\mathcal{G}$, the strategy is that we start at a region where a representation of $\mathcal{G}$ is defined on physical sheet and consistent with perturbation theory result Eq.(\ref{feynmantriangle}), then,    $\mathcal{G}$ is continued to other regions by using   perturbation theory result Eq.(\ref{feynmantriangle}) as a reference.

(1) The dispersion representation of a triangle diagram for \mbox{$t>4 m^{2}$} has been discussed in \cite{Bronzan:1963xn},
\begin{align}
   \mathcal{G}(s,t)  =& \frac{1}{\pi} \int_{4 m^{2}}^{\infty} d s' \frac{1}{s'-s  } \left [  \frac{1 }{ k(s')} \int_{t_{-}(s')}^{t_{+}(s')}  \!\!\!\!\!\!\!\!\! \!\!\! (\Gamma)  \  \ d t'       \frac{1}{t-t'}   \right ] , \label{disptrikernel} \nonumber \\
& \quad \quad \quad \quad  \quad \quad \quad \quad  \mbox{for} \ \ t> 4 m^{2}.
\end{align} 

(2) The Pasquier inversion representation of a triangle diagram for \mbox{$s < (M-m)^{2}$} is given by \cite{Guo:2014vya},
 \begin{align}\label{pastrianglediagram} 
 \mathcal{G}(s,t)   = &   \frac{1}{\pi} \int_{- \infty}^{(M-m)^{2}} \!\!   \frac{d t' }{t-t'}   \left [ \theta(t') \Delta(s, t' ) -\theta(-t') \Sigma(s, t' )  \right ]    , \nonumber \\
& \quad \quad \quad \quad  \quad \quad    \quad \quad       \mbox{for} \ \ s < (M-m)^{2}.
\end{align}

 \begin{figure}
\includegraphics[width=0.45\textwidth]{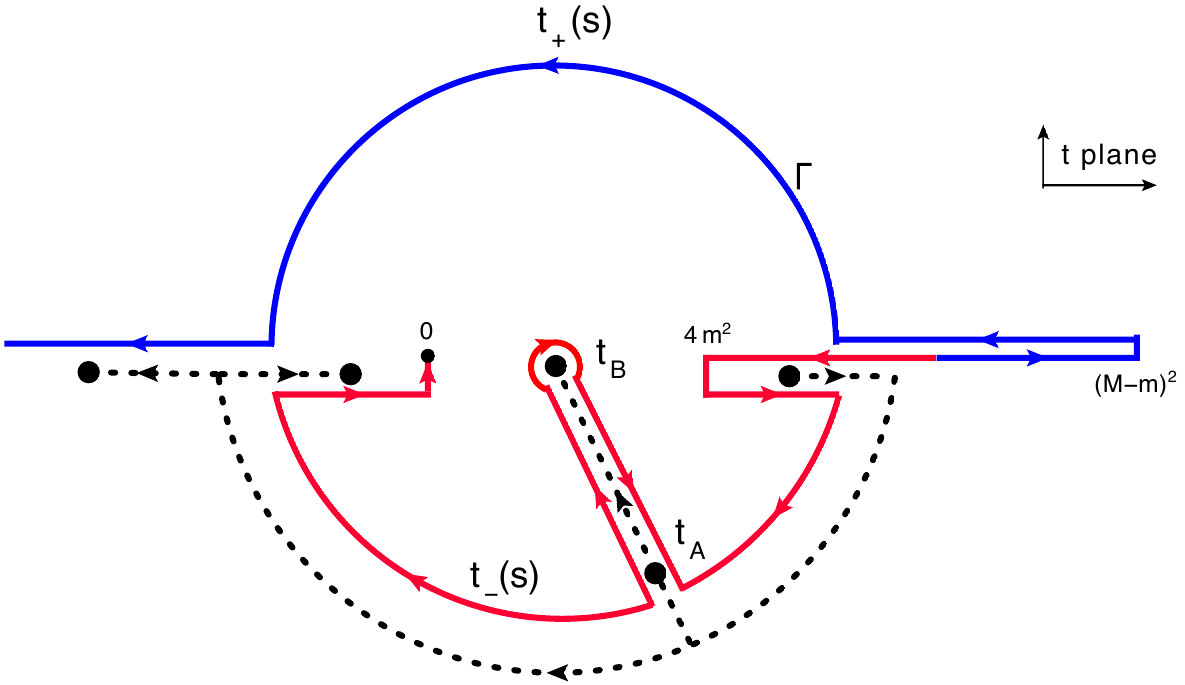}
\caption{   Analytic continuation of  a function  in $t$ of type,  $\mathcal{G}(s,t)$ in Eq.(\ref{tridispexchange}), is followed by the motion of $t$ (black dashed curve).   The physical sheet  of $\mathcal{G}(s,t)$ in $t$ is defined in outside  space of  $\Gamma$. $t\in[0,4m^{2}]$ inside of $\Gamma$ region can be reached by crossing $\Gamma$, then a discontinuity   has to be picked up to keep $\mathcal{G}(s,t)$  on physical sheet.
\label{disptriangle}}
\end{figure}

\subsection{Analytic continuation of dispersion representation of triangle diagram}

We first perform analytic continuation of dispersion representation of $\mathcal{G}$ in Eq.(\ref{disptrikernel}). Note that the overlapping   region for  both dispersion representation and Pasquier inversion representation of $\mathcal{G}$   on physical sheet is \mbox{$t \in [4 m^{2}, \infty]$} and  \mbox{$s \in[-\infty,(M-m)^{2}]$}.   As described in Appendix~\ref{pasquier}, 
 exchanging  order of double integrals encounters no extra singularities in this region,  so, we start from here and rewrite Eq.(\ref{disptrikernel}) to, see Eq.(\ref{doubleintef}-\ref{inverseintef}),
\begin{align}
  \mathcal{G}(s,t)  =& \frac{1}{\pi}  \int_{\Gamma} d t'    \frac{1}{t-t'}  \left [ \int_{s_{\Gamma} (t')}^{\infty}  \!\!\!\!\!\!\!\!\!\! (C)  \   d s'   \frac{1}{s'-s} \frac{N(s') }{U(s')} \right ]    ,   \nonumber \\
& \quad \quad \quad \quad          \mbox{for} \ \ t> 4 m^{2}, s <(M-m)^{2}. \label{tridispexchange}
\end{align} 
The cut in $t$ generated by contour $\Gamma$ is now explicitly given by \mbox{$ \int_{\Gamma} d t' /(t-t')$},  
we start with \mbox{$t$}  running along the black wiggle line in Fig.\ref{fig:2}, where $  \mathcal{G}$ is defined on physical sheet.  As long as the motion of $t$ in complex plane does not interfere with the contour $\Gamma$, $  \mathcal{G}$ remains on physical sheet, thus, Eq.(\ref{disptrikernel}) still holds for \mbox{$t<0$}, see the motion of $t$ represented by black dashed curve in Fig.\ref{disptriangle}. However, when $t$ is moved to cross contour $\Gamma$, the contour $\Gamma$ has to be deformed to keep $  \mathcal{G}$    on physical sheet. To reach region \mbox{$t\in[0,4m^{2}]$}, we can  first move $t$ to     $t_{A}$ which  is a point sits right inside circle of $\Gamma$, see  Fig.\ref{disptriangle}. Thus, the deformation of $\Gamma$ leads to
\begin{align}
  \mathcal{G}(s,t_{A})  =& \frac{1}{\pi}  \int_{\Gamma} d t'    \frac{1}{t_{A}-t'}  \left [ \int_{s_{\Gamma} (t')}^{\infty}  \!\!\!\!\!\!\!\!\!\! (C)  \   d s'   \frac{1}{s'-s} \frac{N(s') }{U(s')} \right ]    \nonumber \\
  +& 2 i \int_{s_{\Gamma} (t_{A})}^{\infty}  \!\!\!\!\!\!\!\!\!\!\! (C)  \   d s'   \frac{1}{s'-s} \frac{N(s') }{U(s')} , \label{disprepGcomplex}  \nonumber \\
& \quad \quad \quad \quad     \quad \quad        \mbox{for} \ \  s <(M-m)^{2}.
\end{align} 
When $t_{A}$ is moved to \mbox{$t_{B} \in [0,4 m^{2}]$} on real axis, contour $\Gamma$ in second piece on the right hand side of Eq.(\ref{disprepGcomplex}) is dragged by the motion of $t$ to collapse onto real axis, see in Fig.\ref{collapsepath}, accordingly, $C$ has to be opened up to $C'$. Thus, we obtain,
\begin{align}
    \mathcal{G}(s,t_{B})  =&\frac{1}{\pi} \int_{4 m^{2}}^{\infty} d s' \frac{1}{s'-s  } \left [  \frac{1 }{ k(s')} \int_{t_{-}(s')}^{t_{+}(s')}  \!\!\!\!\!\!\!\!\! \!\!\! (\Gamma)  \  \ d t'       \frac{1}{t_{B}-t'}   \right ]   \nonumber \\
  +& 2 i \int_{s_{-} (t_{B})}^{\infty}  \!\!\!\!\!\!\!\!\!\!\! (C')  \   d s'   \frac{1}{s'-s} \frac{N(s') }{U(s')} ,\nonumber \\
& \quad \quad \quad \quad     \quad \quad    \quad         \mbox{for} \ \  s <(M-m)^{2}. \label{disprepGs}
\end{align} 
So continuation in $t$ is complete. Next, we need to continue $s$ to the region \mbox{$s\in[(M-m)^{2}, \infty]$}, the continuation of first term on the right hand side of Eq.(\ref{disprepGs}) shows no difficulty and encounters no extra singularities. However, as we can see in Fig.\ref{pasqtriangle}, $s$ on real axis is divided by contour $C'$ into three sections, thus, for \mbox{$s\in[(M-m)^{2}, (M+m)^{2}]$}, a pole contribution, $-4\pi/U(s)$, is picked up by second term on the right hand side of Eq.(\ref{disprepGs}). In the end,  analytic continuation of dispersion representation of $\mathcal{G}$ is given by
\begin{align}
   \mathcal{G}(s,t)  =& \frac{1}{\pi} \int_{4 m^{2}}^{\infty} d s' \frac{1}{s'-s  }  \left [\frac{1 }{ k(s')} \int_{t_{-}(s')}^{t_{+}(s')}  \!\!\!\!\!\!\!\!\! \!\!\!\! (\Gamma)  \  \ d t'       \frac{1}{t-t'} \right ]  \nonumber \\
+& 2 i  \theta(t) \theta(4 m^{2} -t)  \bigg [ \int_{s_{-} (t)}^{  s_{+} (\infty)}  \!\!\!\!\!\!\!\!\ \!\!\! \!\!\!\! \!\! (C') \   \ d s'  \frac{1}{s'-s }  \frac{ 1 }{  U(s')  } \nonumber \\
&  \   + \theta \left (s-(M-m)^{2} \right ) \theta \left ((M+m)^{2}-s \right ) \frac{2\pi i}{U(s)} \bigg ] , \nonumber \\
& \quad \quad \quad \quad  \quad \quad \quad    \mbox{for} \ \ (s,t) \in [-\infty, \infty]. \label{fulldisptriangle}  
\end{align}

\subsection{Analytic continuation of Pasquier inversion representation of triangle diagram}

For the analytic continuation of Eq.(\ref{pastrianglediagram}), similarly, we start from region \mbox{$t \in [4 m^{2}, \infty], s \in[-\infty,(M-m)^{2}]$}. % in which  exchanging  order of double integrals encounters no extra singularities. 
We first write Eq.(\ref{pastrianglediagram}) to, see Eq.(\ref{collapseintef}-\ref{pasqinvert}),
\begin{align} 
 \mathcal{G}(s,t)   = &   \frac{1}{\pi} \int_{\Gamma'}     \frac{d t' }{t-t'}  \int_{s_{\Gamma'}(t)}^{ \infty}  \!\!\!\!\!\!\!\!\!\!  (C') \ d s'   \frac{1}{s'-s }  \frac{ 1 }{  U(s')  }    , \nonumber \\
& \quad \quad  \quad \quad       \mbox{for} \ \ t> 4 m^{2}, s < (M-m)^{2}.
\end{align} 
By    exchanging the order of two integrals, we obtain,
 \begin{align} 
&  \mathcal{G}(s,t)  =\frac{1}{\pi} \int_{C'}   d s'   \frac{1}{s'-s }  \frac{ 1 }{  U(s')  }    \int_{0}^{ t_{C'}(s')}  \!\!\!\!\!\!\!\!\!\!\!\!\!  (\Gamma') \    \frac{d t' }{t-t'}  ,  \nonumber \\
& \quad \quad \quad \quad      \quad \quad  \quad \quad   \mbox{for} \ \ t> 4 m^{2}, s < (M-m)^{2}. \label{pasqrepgexchange}
\end{align} 
As we can see in Eq.(\ref{pasqrepgexchange}) and also described previously in section \ref{analyticpasquier},  $s$ plane is divided by contour $C'$. Only for the region \mbox{$s\in[(M-m)^{2},(M+m)^{2}]$}, $ \mathcal{G}$ need to pick up an extra term $  2i/   U(s)      \int_{0}^{ t_{C}(s)}   (\Gamma)   d t' /(t-t')$ to stay on physical sheet,  thus, the analytic continuation in $s$  leads to,
 \begin{align} 
&  \mathcal{G}(s,t)  =  \frac{1}{\pi} \int_{- \infty}^{(M-m)^{2}} \!\!  \frac{d t'  }{t-t'}    \left [ \theta(t') \Delta(s, t' ) -\theta(-t') \Sigma(s, t' )  \right ]  \nonumber \\
& + \theta \left (s-(M-m)^{2} \right ) \theta \left ((M+m)^{2}-s \right )   \frac{ 2i }{  U(s)  }    \int_{0}^{ t_{C}(s)}  \!\!\!\!\!\!\!\!\!\!\!\!\!  (\Gamma) \    \frac{d t' }{t-t'}, \nonumber \\
& \quad \quad \quad \quad      \quad \quad  \quad \quad   \mbox{for} \ \ t> 4 m^{2}, s \in [-\infty, \infty]. \label{pasqrepgt}
\end{align} 
Next, we continue $t$ to below $4m^{2}$, again, the first term on the right hand side of Eq.(\ref{pasqrepgt}) shows no difficulty of continuation and remains the same.  From Fig.\ref{disptriangle}, we can see, $t$ plane is divided by contour $\Gamma$, thus, only   second term on the right hand side of Eq.(\ref{pasqrepgt}) for $t\in[0,4m^{2}]$ need to pick up a pole contribution, \mbox{$ -4\pi /U(s)$}, 
to stay on physical sheet. In the end, analytic continuation of Pasquier inversion representation of $\mathcal{G}$ is given by,
 \begin{align}\label{fullpastrianglediagram} 
 \mathcal{G}(s,t)    
& =    \frac{1}{\pi} \int_{- \infty}^{(M-m)^{2}} \!\!  \frac{d t'  }{t-t'}    \left [ \theta(t') \Delta(s, t' ) -\theta(-t') \Sigma(s, t' )  \right ]    \nonumber \\
&+  2 i \theta \left (s-(M-m)^{2} \right ) \theta \left ((M+m)^{2}-s \right ) \nonumber \\
&  \quad  \quad   \times \left [ \frac{1 }{ U(s)} \int_{ 0}^{t_{+}(s)}  \!\!\!\!\!\!\!\!\! \!\! (\Gamma)  \       \frac{d t' }{t-t'} +  \theta(t) \theta(4 m^{2} -t) \frac{2\pi i}{U(s)} \right ], \nonumber \\
& \quad \quad \quad \quad  \quad \quad    \quad \quad          \mbox{for} \ \ (s,t) \in [- \infty, \infty].
\end{align}

%%%%%%%%%%%%%%%%%%%%%%%%%%%%%%%%%%%%%%%%%%%%%%%%%%%%%%%%%%%%%%%%%%%%%%%%%%%%%%%%%%%%%%%
%%%%%%%%%%%%%%%%%%%%%%%%%%%%%%%%%%%%%%%%%%%%%%%%%%%%%%%%%%%%%%%%%%%%%%%%%%%%%%%%%%%%%%%
%%%%%%%%%%%%%%%%%%%%%%%%%%%%%%%%%%%%%%%%%%%%%%%%%%%%%%%%%%%%%%%%%%%%%%%%%%%%%%%%%%%%%%%

\end{document}